\documentclass[lettersize,journal]{IEEEtran}
\usepackage{booktabs}   
\usepackage{subcaption} 
\usepackage{algorithmic}
\usepackage{graphicx}
\usepackage{xcolor}
\usepackage{graphicx}
\usepackage[linesnumbered, ruled, vlined]{algorithm2e}
\usepackage{makecell}
\usepackage{pifont}
\usepackage{enumitem}
\usepackage{cuted}
\usepackage{multirow}

\begin{document}

\title{Cascaded Prediction and Asynchronous Execution of Iterative Algorithms on Heterogeneous Platforms}

\author{Jianhua Gao, Bingjie Liu, Yizhuo Wang, Weixing Ji\textsuperscript{*}, Hua Huang
\thanks{\textsuperscript{*}Corresponding author.}
\thanks{Jianhua Gao, Weixing Ji, and Hua Huang are with the School of Artificial Intelligence, Beijing Normal University, Beijing 100875, China (e-mail: gaojh@bnu.edu.cn, jwx@bnu.edu.cn, huahuang@bnu.edu.cn). Bingjie Liu and Yizhuo Wang are with the School of Computer Science and Technology, Beijing Institute of Technology, Beijing 100086, China (e-mail: liubj@bit.edu.cn, frankwyz@bit.edu.cn).}
}

\markboth{Journal of \LaTeX\ Class Files,~Vol.~14, No.~8, August~2021}%
{Shell \MakeLowercase{\textit{et al.}}: A Sample Article Using IEEEtran.cls for IEEE Journals}


\maketitle

\begin{abstract}
Owing to the diverse scales and varying distributions of sparse matrices arising from practical problems, a multitude of choices are present in the design and implementation of sparse matrix-vector multiplication (SpMV). Researchers have proposed many machine learning-based optimization methods for SpMV. However, these efforts only support one area of sparse matrix format selection, SpMV algorithm selection, or parameter configuration, and rarely consider a large amount of time overhead associated with feature extraction, model inference, and compression format conversion. This paper introduces a machine learning-based cascaded prediction method for SpMV computations that spans various computing stages and hierarchies. Besides, an asynchronous and concurrent computing model has been designed and implemented for runtime model prediction and iterative algorithm solving on heterogeneous computing platforms. It not only offers comprehensive support for the iterative algorithm-solving process leveraging machine learning technology, but also effectively mitigates the preprocessing overheads. Experimental results demonstrate that the cascaded prediction introduced in this paper accelerates SpMV by 1.33x on average, and the iterative algorithm, enhanced by cascaded prediction and asynchronous execution, optimizes by 2.55x on average.
\end{abstract}


\section{Introduction}
Conjugate gradient (CG) and global minimum residual (GMRES), both grounded in the Krylov subspace, are the mainstream iterative algorithms for solving large-scale sparse linear equations. These algorithms include sparse matrix-vector multiplication (SpMV), vector inner product, and other computational kernels. Among them, SpMV comprises the largest proportion of the overall running time of the two iterative algorithms \cite{TPDS-largeSpMV}, thereby garnering extensive attention from researchers over the past two decades \cite{SPMV-survey}.

Typically, sparse matrices from real-world engineering problems are rich in zeros, necessitating storage in a compressed format. The irregular distribution of non-zeros in sparse matrices has attracted different compression format designs. The considerable research effort is focused on designing new sparse matrix compression formats and their associated SpMV algorithms \cite{SPMV-survey}.
However, according to existing studies, no compression format has been found that consistently delivers good performance across all sparse matrices \cite{Akrem-BestSF}, and there is significant variation in the performance of different SpMV algorithms designed for the same sparse format. The most notable example is the implementation of SpMV in the CSR format.
Algorithms based on CSR employ diverse strategies such as sparse matrix partitioning, data cache optimization, and load balancing, resulting in significant disparities in SpMV's computational efficiency across different sparse matrices. Moreover, even within a single algorithm, issues of parameter setting arise, with optimal parameter configurations differing for various sparse matrices. For instance, Gao et al. \cite{jpdc/GaoJLWS24} examined the thread configuration issue of the CSR-Vector algorithm \cite{bell2008efficient,baskaran2008optimizing} in CUSP \cite{CUSP}. The SpMV using the optimal thread configuration achieved an average speedup of 3x and a maximum speedup of 20x compared with the SpMV with default settings.
 
As machine learning (ML) technology becomes more widespread, researchers address these issues by gathering performance data and training machine learning models. These efforts can be classified mainly into three categories: the optimal compression format prediction \cite{SMAT2013}\cite{Akrem2016ICPP_SVM}\cite{DBLP:conf/ics/SedaghatiMPPS15}\cite{Akrem-BestSF}\cite{ShenXipeng2018PPoPP_bridgingGap}, the optimal SpMV algorithm prediction \cite{2017-IPDPSW-autoTuning}\cite{DBLP:journals/ieicet/CuiHKT18}\cite{ijhpca/DufrechouEQ21}, and the optimal parameter setting prediction \cite{2017-IPDPSW-autoTuning}\cite{usman2019zaki}\cite{DBLP:journals/access/UsmanMKA19}\cite{ahmed2022aaqal}\cite{jpdc/GaoJLWS24}. Although these methods have improved SpMV performance to some extent, they still face the following challenges:
(1) \textbf{Current ML-based optimization efforts only cover the prediction of a single area} within the compressed formats, SpMV algorithms, or parameter settings. The search space is so small that there is still a gap from the best SpMV performance.
(2) \textbf{The ML-based SpMV and iterative algorithm optimization introduce significant preprocessing overhead}, which offsets the performance benefits from ML-based optimization.

Addressing the aforementioned challenges, this paper introduces an ML-based cascade prediction approach for typical iterative algorithms like CG and GMRES, targeting heterogeneous platforms. This optimization spans three key areas: selecting sparse matrix compression formats, SpMV algorithms, and parameter configurations. Building on this, an asynchronous execution model is proposed, integrating model inference with iterative computation. This model conceals the costs associated with model inference and matrix preprocessing, fully leverages the heterogeneous computing resources of CPU-GPU, and improves the operational efficiency of iterative algorithms on these platforms.

This paper's main contributions include:
\setlist[itemize]{leftmargin=0.5cm, itemindent=0pt}
\begin{itemize}
\item We propose a lightweight cascading prediction framework for ML-based SpMV to mitigate the suboptimal performance caused by single area prediction. 
\item We propose an asynchronous execution model for iterative algorithms on heterogeneous platforms to hide the preprocessing overheads.
\item We take the GMRES algorithm as an example to prove the efficiency of our method using 22 sparse matrices.
\end{itemize}

\section{Motivation}
\subsection{Single Area Prediction}
We considered three libraries that support sparse matrix computations: cuSPARSE \cite{nvidia2023}, CUSP \cite{CUSP}, and MAGMA \cite{magma-library}, and tested the performance of SpMV in cuSPARSE based on COO and CSR formats, in CUSP based on COO, CSR, DIA, HYB, and ELL formats, and the CSR-Vector algorithm in CUSP with different thread configurations. We denote the method with the shortest SpMV running time among all these implementations as the optimal SpMV configuration. Taking different sparse matrix compression formats in cuSPARSE as an example, Table \ref{tab:motivate-1} compares the SpMV running time using the optimal compression format in cuSPARSE and the running time of the optimal SpMV configuration on two sparse matrices. We can observe that the optimal SpMV configuration on both matrices is the CSR-Vector algorithm in CUSP. Selecting the fastest format in cuSPARSE, COO format, still falls short of the performance of the optimal SpMV configuration in the selection space explored in this paper. 

\begin{table}[!htb]
    \centering
    \caption{Comparison of SpMV running time (in milliseconds) using the different compression formats in cuSPARSE and the SpMV running time using optimal SpMV configuration.}
    \scalebox{0.8}{
    \begin{tabular}{c|>{\centering\arraybackslash}m{1.7cm}|>{\centering\arraybackslash}m{1.7cm}|>{\centering\arraybackslash}m{1.7cm}>{\centering\arraybackslash}m{2.2cm}}
        \toprule
        \multirow{2}*{\textbf{Matrix}} & \multirow{2}{1.7cm}{\textbf{cuSPARSE-COO}} & \multirow{2}{1.7cm}{\textbf{cuSPARSE-CSR}} & \multicolumn{2}{c}{\textbf{Optimal SpMV}} \\ \cmidrule{4-5}
        & & & \textbf{Time} & \textbf{Configuration} \\ \midrule
        n4c5-b4 & 0.0067 & 0.0086 & 0.0031 & CUSP-CSR-2 \\ 
        bcsstk10 & 0.0066 & 0.0080 & 0.0033 & CUSP-CSR-32 \\
        \bottomrule
    \end{tabular}}
    \label{tab:motivate-1}
\end{table}

Similarly, Table \ref{tab:motivate-2} compares the running time of the different CSR-based SpMV with the running time using the optimal SpMV configuration. On the two example matrices, the optimal SpMV is the COO-based SpMV in cuSPARSE, achieving speedups of 1.31x and 1.30x to the optimal CSR-based SpMV, respectively. So there is still a performance gap between the predicted optimal CSR-based SpMV algorithm and the optimal SpMV configuration. In addition, Table \ref{tab:motivate-3} compares the running time of SpMV using the different parameter settings and the SpMV using the optimal SpMV configuration. We can observe that even with the optimal parameter configuration, there is still a significant performance gap compared with the optimal SpMV configuration.

\begin{table}[!htb]
    \centering
    \caption{Running time (ms) comparison of CSR-based SpMV and the SpMV using the optimal SpMV configuration.}
    \scalebox{0.8}{
    \begin{tabular}{>{\centering\arraybackslash}m{1.2cm}|>{\centering\arraybackslash}m{1cm}|>{\centering\arraybackslash}m{1.7cm}|>{\centering\arraybackslash}m{1.5cm}|>{\centering\arraybackslash}m{1.5cm}>{\centering\arraybackslash}m{1.8cm}}
        \toprule
        \multirow{2}{1.2cm}{\textbf{Matrix}} & \multirow{2}{1cm}{\textbf{CUSP-CSR}} & \multirow{2}{1.7cm}{\textbf{cuSPARSE-CSR}} & \multirow{2}{1.5cm}{\textbf{MAGMA-CSR}} & \multicolumn{2}{c}{\textbf{Optimal SpMV}} \\ \cmidrule{5-6}
        & & & & \textbf{Time} & \textbf{Configuration} \\ \midrule
        iprob & 0.1543 & 0.0088 & 0.2130 & 0.0067 & cuSPARSE-COO \\ \midrule
        adder\_tr\newline ans\_01 & 0.0238 & 0.0086 & 0.0811 & 0.0066 & cuSPARSE-COO\\
        \bottomrule
    \end{tabular}}
    \label{tab:motivate-2}
\end{table}

\begin{table}[!htb]
    \centering
    \caption{Running time (ms) comparison of SpMV using the different parameter settings and using the optimal SpMV configuration.}
    \scalebox{0.8}{
    \begin{tabular}{
    >{\centering\arraybackslash}m{1.2cm}|
    >{\centering\arraybackslash}m{0.7cm}
    >{\centering\arraybackslash}m{0.7cm}
    >{\centering\arraybackslash}m{0.7cm}
    >{\centering\arraybackslash}m{0.8cm}
    >{\centering\arraybackslash}m{0.8cm}|
    >{\centering\arraybackslash}m{0.8cm}
    >{\centering\arraybackslash}m{1.7cm}}
        \toprule
        \multirow{2}{1.2cm}{\textbf{Matrix}} & \multicolumn{5}{c|}{\textbf{Threads per Vector (TpV)}} & \multicolumn{2}{c}{\textbf{Optimal SpMV}}\\ \cmidrule{2-8}
        & \textbf{2} & \textbf{4} & \textbf{8} & \textbf{16} & \textbf{32} & \textbf{Time} & \textbf{Configuration} \\ \midrule
        iprob & 0.1531 & 0.0759 & 0.0372 & 0.0206 & 0.0124 & 0.0067 & cuSPARSE-COO\\ \midrule
        adder\_tr\newline ans\_01 & 0.0667 & 0.0337 & 0.0194 & 0.0112 & 0.0081 & 0.0066 & cuSPARSE-COO\\
        \bottomrule
    \end{tabular}}
    \label{tab:motivate-3}
\end{table}		

The above analysis indicates that predicting the optimal choice from a single area is insufficient. Choosing the optimal SpMV configuration from multiple dimensions is expected to enhance performance further.

\begin{figure*}[!htbp]
    \centering
    \includegraphics[width=0.8\linewidth]{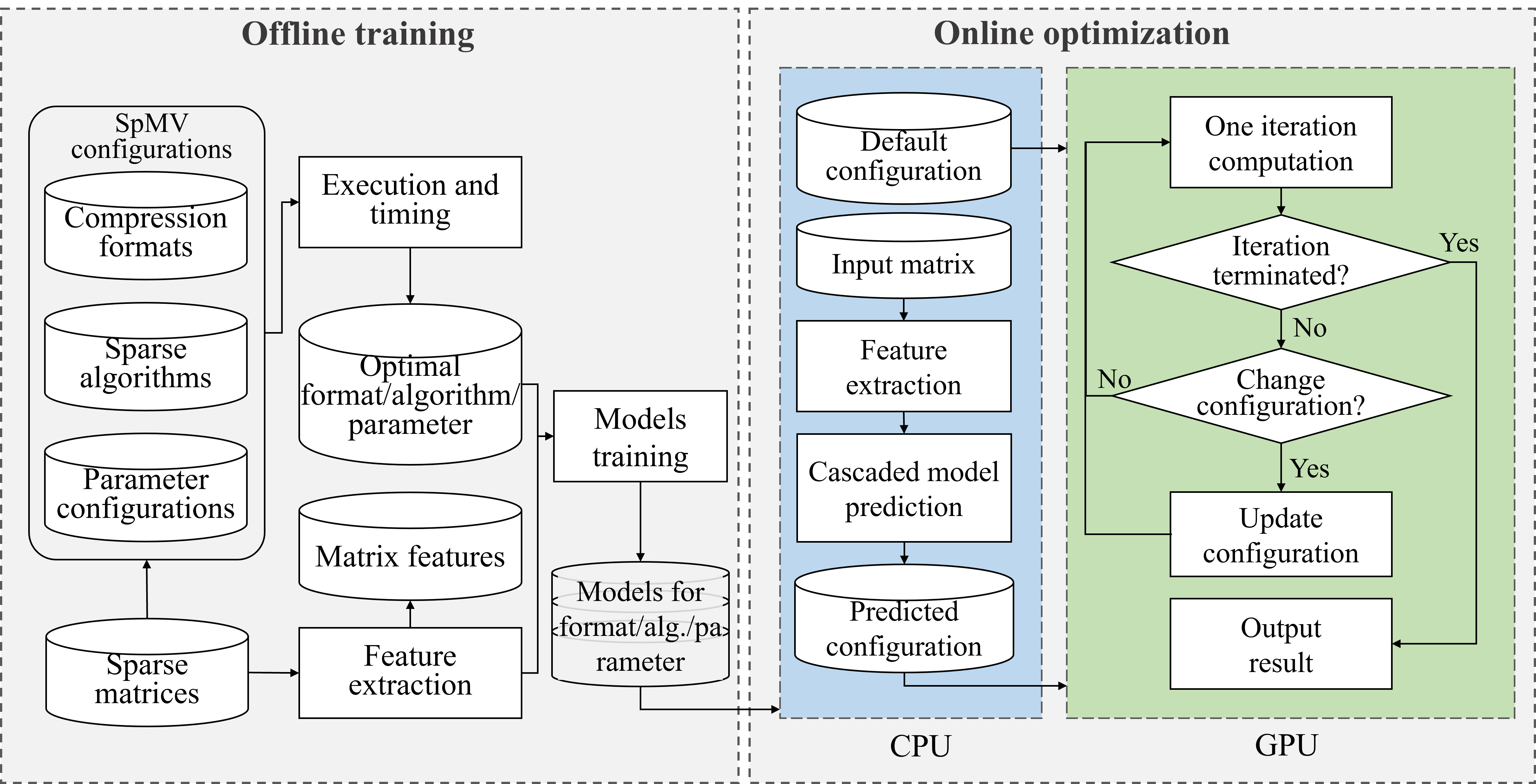}
    \caption{Overview for cascaded prediction and asynchronous execution of iterative algorithms on CPU-GPU heterogeneous platforms.}
    \label{fig:ml-spmv-workflow}
\end{figure*}

\subsection{Preprocessing Cost}
The preprocessing overhead of current ML-based methods mainly includes sparse matrix feature extraction, format conversion, and model prediction. The statistical results from Swann et al. \cite{cgo/SwannOSM24} show that when the sparse matrix size is less than 100,000, the overhead of matrix feature extraction is significant compared with the cost of a single SpMV kernel. When the matrix size is greater than 100,000, the feature extraction overhead is usually less than the running time of a single SpMV kernel. This indicates that when the matrix size is small, the feature extraction overhead needs to be carefully considered; when the matrix size is large, if the feature extraction overhead can be avoided, it is expected to further improve the SpMV performance. In terms of format conversion overhead, the investigation from Shen et al. \cite{Shenxipeng2020TPDS_overheadConscious} shows that the overheads of converting sparse matrices from CSR to COO, DIA, ELL, HYB, BSR, and CSR5 formats, normalized by the time of a single CSR-based SpMV, are 9, 270, 102, 147, 47, and 26, respectively. It also means that format conversion overhead can only be amortized when there are enough SpMV calls in an iterative algorithm. In terms of model prediction, we tested the inference overhead of a Python model trained in this paper, which is about 1-2 ms. However, the running time of a single SpMV kernel may be several microseconds, tens of microseconds, or hundreds of microseconds. This means that on some sparse matrices, the time overhead of model inference is significant. 


\section{Methodology}
\subsection{Overview}
To address the issues of suboptimal performance in existing work that only considers single area prediction and the SpMV performance benefits reduction due to the additional preprocessing overhead, this paper introduces an optimization method, combining cascaded prediction with asynchronous execution, for iterative algorithms based on CPU-GPU heterogeneous platforms. Figure \ref{fig:ml-spmv-workflow} presents the overall framework of this method. The framework has two stages: offline training and online optimization. During the offline model training, the features of each sparse matrix, as well as the execution time of each matrix across various SpMV configurations, are first collected.
The configuration with the shortest execution time is labeled and combined with the matrix features to generate datasets. ML models are then trained for different SpMV configurations. 

During the online optimization, given an iterative solver and the default SpMV configuration, we apply cascaded prediction and asynchronous execution to optimize the iterative algorithm. 
The CPU is tasked with extracting features from sparse matrices and predicting the optimal SpMV configuration using the cascaded approach, which infers compression format, SpMV algorithm, and parameter configuration in a cascaded way. Concurrently, the iterative solver executes on the GPU side with a default SpMV configuration. If the current iteration has not converged or reached the maximum iterations, then check the model's predicted results from the CPU side. 
If the inference of any model has been completed, compare it with the currently used SpMV configuration to decide whether to update the SpMV configuration; Otherwise, the iteration continues.

By cascading predictions from multiple models, we explore a broader range of options, addressing the suboptimal performance issues associated with predictions for a single configuration. With the asynchronous execution of iterative algorithms and model predictions, we conceal the preprocessing overhead and improve the utilization of computing resources on heterogeneous platforms.

\subsection{Offline Model Trainning}
To develop ML models for cascaded prediction, we need to build a dataset for model training. We extract sparse matrix features and gather performance data for SpMV across various configurations, labeling the configuration with the shortest running time. Subsequently, we use AutoML to train machine-learning models for different SpMV configurations. We next provide a detailed and in-depth introduction of the research methods used in each stage.

\subsubsection{Features Extraction}
This paper selects a universal feature set to predict the optimal SpMV across three SpMV configurations. Table \ref{tab:feature-info} presents their meanings, computational complexity, and calculation formulas. In which, $nrows$, $ncols$, and $nnz$ can be directly obtained. Then, $density$ and $mean$ can be calculated easily using the three features, thus both having a time complexity of $O(1)$. Moreover, $sd$, $max$, and $min$ are associated with the distribution of non-zeros across matrix rows. These can be computed by traversing the row pointer array of CSR, with the computational complexity being decided by the sparse matrix's rows. Additionally, $cov$, $maxavg$, and $fill$ rely on other features, thereby taking the computational cost of $O(1)$. $Distavg$, $clusteravg$, $ndiag$, and $diagfill$ offer more in-depth observations of the matrix. Their calculation requires traversing all non-zeros, with the time cost reaching the $O(nnz)$. Similarly, $diagfill$ is calculated using $ndiag$, thus having a computational complexity of $O(1)$. In summary, the chosen 15 features characterize the non-zeros' distribution from various aspects, enabling ML models to better model the relationship between matrices and optimal SpMV configurations.

\begin{table}[!htbp]
    \centering
    \caption{Sparse matrix feature set.}
    \scalebox{0.75}{
    \begin{tabular}{m{1.6cm}m{2.8cm}cm{3.6cm}}
        \toprule
        \textbf{Feature} & \multicolumn{1}{c}{\textbf{Meaning}} & \textbf{Complexity} & \multicolumn{1}{c}{\textbf{Formula}} \\
        \midrule
        $nrows$, $ncols$, $nnz$ & Rows, columns, and non-zeros & $O(1)$ & / \\ \midrule
        $density$ & Density of non-zeros & $O(1)$ & $nnz/(nrows*ncols)$ \\ \midrule
        $mean$ & Average NNZ per row & $O(1)$ & $\frac{1}{nrows}\sum_{i=1}^{nrows}{r_i}$, $r_i$ denotes the non-zeros in the $i$-th row \\ \midrule
        $sd$ & The standard variance of NNZ per row & $O(nrows)$ & $\sqrt{\frac{1}{nrows}\sum_{i=1}^{nrows}(mean-r_i)^2}$ \\ \midrule
        $cov$ & The covariance of NNZ per row & $O(1)$ & $sd / mean$ \\ \midrule
        $max$,$min$ & Maximum and minimum NNZ per row & $O(nrows)$ & $max_{i=1}^{nrows}r_i$,$min_{i=1}^{nrows}r_i$ \\ \midrule
        $maxavg$ & The difference between $max$ and $mean$ & $O(1)$ & $max - mean$ \\ \midrule
        $distavg$ & Average column width per row & $O(nnz)$ & $\frac{1}{nrows}|c_i^{f}-c_i^{l}|$, $c_i^{f}$ and $c_i^{l}$ denote the column indices of the first and last non-zero element per row \\ \midrule
        $clusteravg$ & The average of the maximum NNZ distributed continuously per row & $O(nnz)$ & $\frac{1}{nrows}\sum_{i=1}^{nrows}cl_{i}$, $cl_{i}$ denotes the maximum NNZ distributed continuously in the $i$-th row \\ \midrule
        $fill$ & Filling rate & $O(1)$ & $\frac{1}{nnz}nrows*max$ \\ \midrule
        $ndiag$ & Diagonals containing at least one non-zero element & $O(nnz)$ & / \\ \midrule
        $diagfill$ & Filling rate of diagonals & $O(1)$ & $\frac{1}{nnz}nrows*ndiag$ \\
        \bottomrule
        \end{tabular}}
    \label{tab:feature-info}
\end{table}

\subsubsection{Dataset construction}
\begin{figure}[!htbp]
    \centering
    \includegraphics[width=0.8\linewidth]{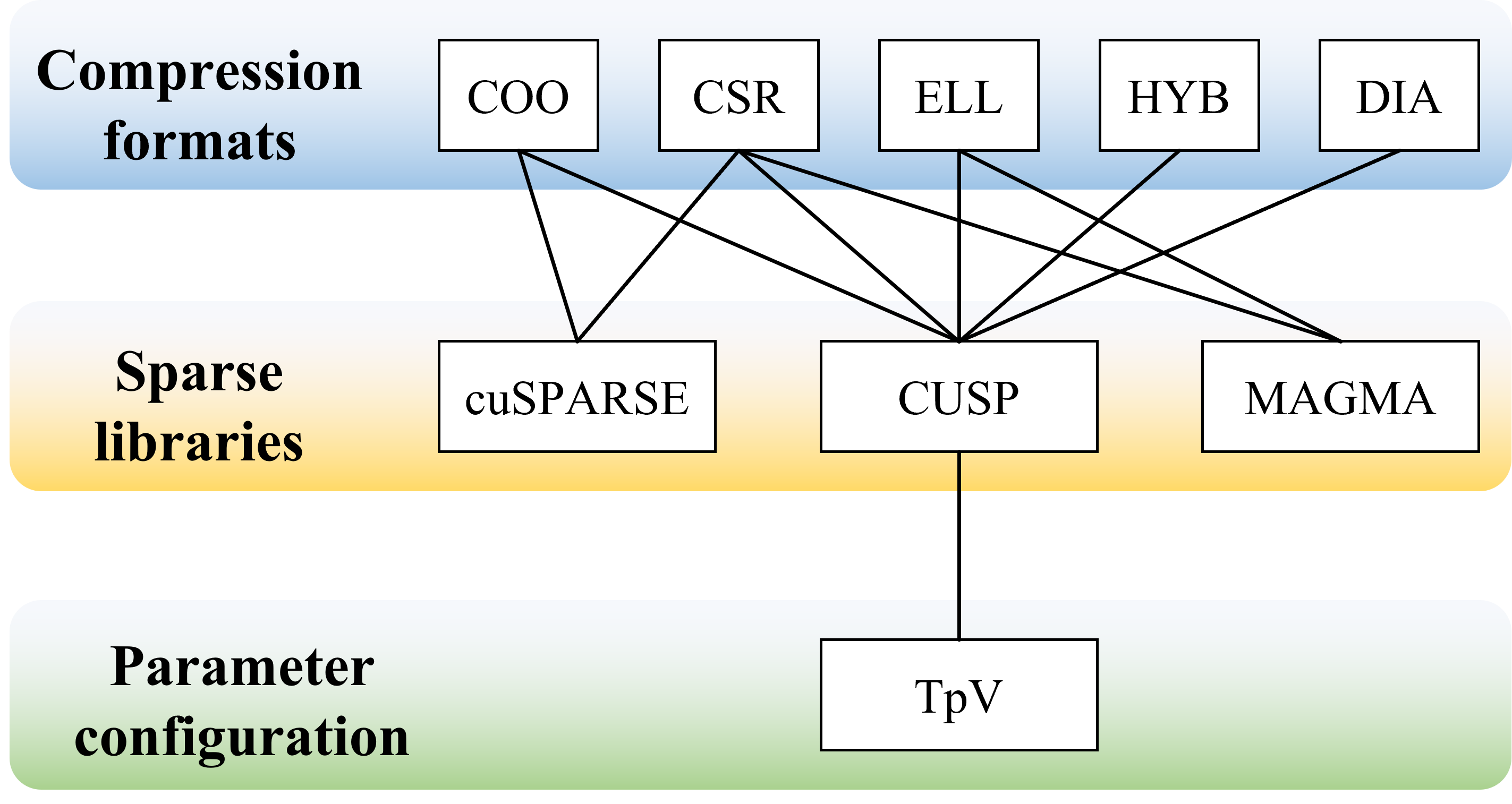}
    \caption{SpMV configuration space explored in this paper.}
    \label{fig:ml-select-space}
\end{figure}

The sparse matrices used for training and testing models are all sourced entirely from the SuiteSparse Matrix Collection \cite{Davis2011_SuiteSparse}. 
It includes a total of 2,893 sparse matrices. Apart from sparse matrices containing complex numbers, and those encountering operational exceptions or failures, the finally constructed dataset includes 2,581 sparse matrices. Besides, we consider three SpMV configurations: sparse matrix compression formats, SpMV algorithms across various computing libraries, and parameter configurations. The compression formats considered in this paper include COO, CSR, ELL, DIA, and HYB. The computing libraries primarily considered are CUSP \cite{CUSP}, cuSPARSE \cite{nvidia2023}, and MAGMA \cite{magma-library}\cite{magma-paper}. The kernel parameter configuration primarily involves the parameter threads per vector ($TpV$) of the CSR-Vector method in CUSP,  with possible values of 2, 4, 8, 16, and 32. Finally, we outline the SpMV configuration space as depicted in Figure \ref{fig:ml-select-space}.

\begin{figure}[!htbp]
    \centering
    \includegraphics[width=0.99\linewidth]{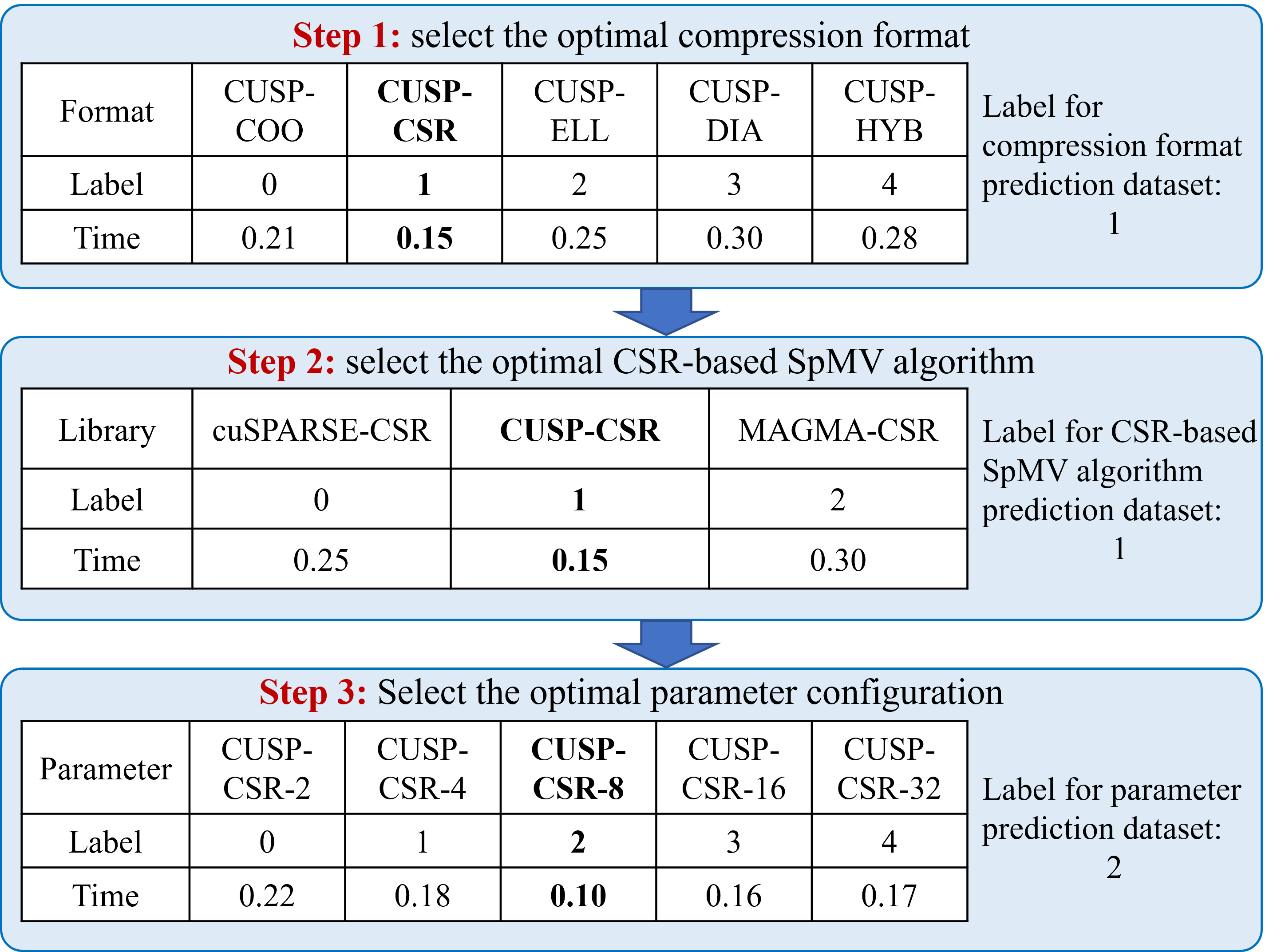}
    \caption{Example of generating labels for different SpMV configuration prediction datasets.}
    \label{fig:label-sample}
\end{figure}

After giving the sparse matrix set and the SpMV configuration selection space, one can begin collecting the SpMV running time for each sparse matrix across various configurations. 
We record the time of 200 SpMV runs and compute the average as the final result. Next, we label each sparse matrix for various SpMV configurations. Usually, the compression format has the greatest impact on SpMV performance, followed by the SpMV algorithm and kernel parameter configurations. Considering the sparse matrix compression formats supported by CUSP are the most comprehensive among the considered three libraries, the optimal sparse matrix compression format is determined by comparing SpMV times across different formats in the CUSP library. 
Subsequently, for each sparse matrix compression format, the running time of SpMV algorithms from different computation libraries is compared, with the library providing the shortest running time serving as the label of the SpMV algorithm prediction dataset under the compression format.
Figure \ref{fig:label-sample} presents an example of generating labels for the compression format prediction dataset, SpMV algorithm prediction dataset, and parameter setting prediction dataset.

Consequently, for three SpMV configurations, we have generated five datasets: one for predicting the optimal sparse matrix compression format, another for choosing the best SpMV algorithm for CSR format, a third for selecting the best SpMV algorithm for ELL format, a fourth for determining the optimal SpMV algorithm for DIA format, and a fifth for predicting the best parameter configuration.

\subsubsection{Model Training}

AutoML technology enables automated model selection and hyperparameter tuning, lowering the barrier for non-experts to use ML techniques to solve practical problems. AutoGluon \cite{nick2019autogluon} is an AutoML framework launched by the Amazon team. It automatically searches the model and hyperparameter spaces, helping users obtain highly accurate machine-learning models on a given dataset. 

\begin{figure}[!htbp]
    \centering
\includegraphics[width=0.7\linewidth]{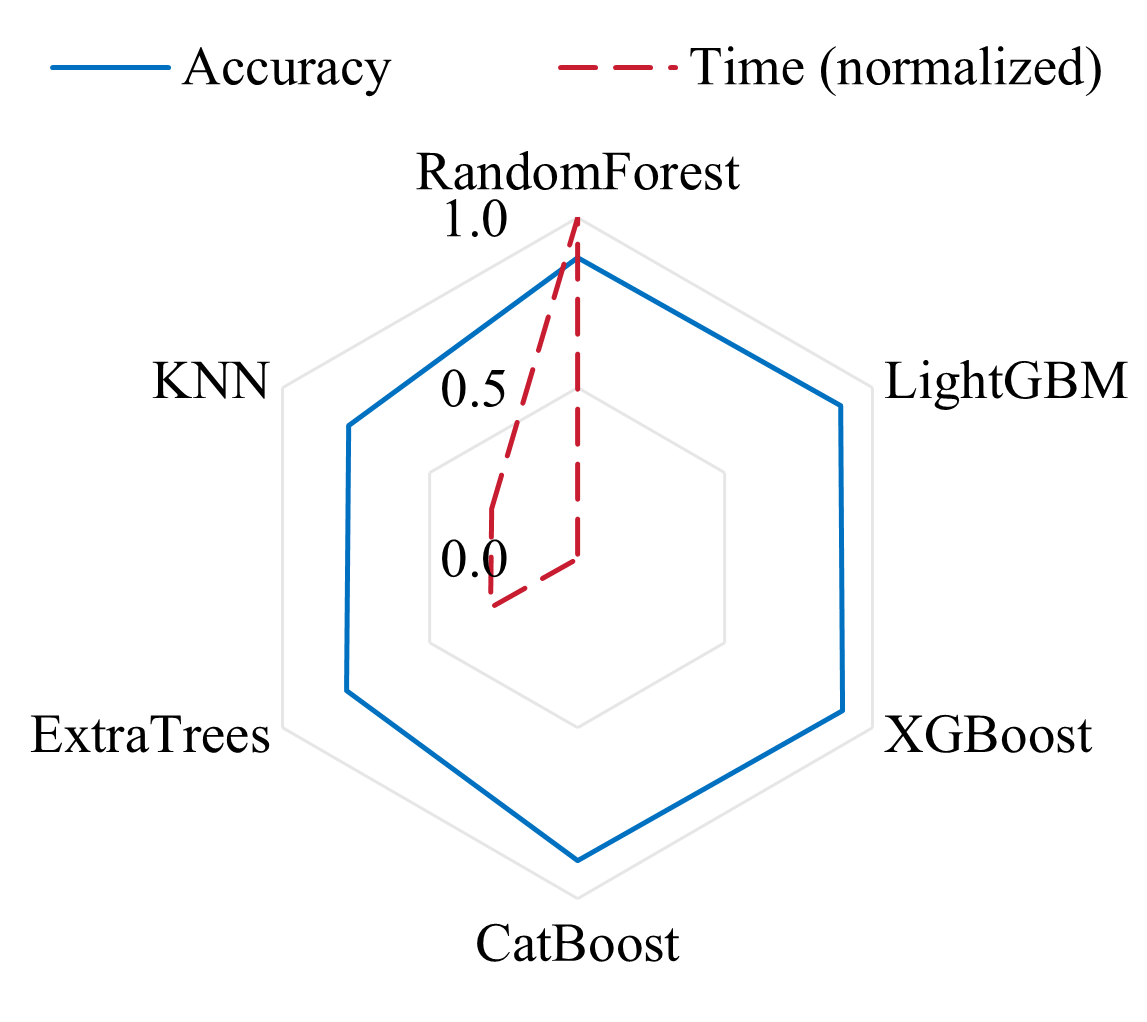}
    \caption{Comparison of prediction accuracy and time (using min-max normalization) of different FORMAT models.}
    \label{fig:fit-radar}
\end{figure}

According to the model's preconditions and prediction objectives, we need to train five models: FORMAT, COO-LIB, CSR-LIB, ELL-LIB, and CSR-CUSP-TPV, respectively. Choosing the optimal SpMV configuration for sparse matrices is a classification problem in supervised learning. The AutoGluon framework provides multiple model choices, including neural networks (NN) \cite{ANN-survey}, LightGBM \cite{ke2017lightgbm}, CatBoost \cite{prokhorenkova2018catboost}, XGBoost \cite{chen2016xgboost}, Random Forest \cite{breiman2001random}, Extra Trees \cite{geurts2006extremely}, and $k$ Nearest Neighbors (KNN) \cite{peterson2009k}. Neural network models have poor interpretability and often require large datasets for training to achieve good generalization performance, which poses a risk of overfitting when the total amount of data is small. Therefore, it is not considered in this work. 
We collected the test accuracy and prediction time of the models trained using other six ML algorithms.
To make the comparison more intuitive, we use a radar figure to display the results of the FORMAT model, as shown in Figure \ref{fig:fit-radar}. The prediction time is processed using the min-max normalization. The prediction accuracy of XGBoost, LightGBM, CatBoost, and RandomForest models all exceeds 88\%. The prediction time of RandomForest is higher than that of other models, while the prediction time costs of LightGBM, CatBoost, and XGBoost models are lower. Considering comprehensively, we select the LightGBM model as the sparse matrix compression format prediction model. The same analysis was conducted on the other four prediction tasks, and the results showed that the best-performing models are all LightGBM. 

\subsection{Online Optimization}
On the one hand, the number of SpMV calls in practical applications is unknown. Taking iterative solving algorithms as an example, the number of SpMV calls depends on the number of iterations of the iterative algorithm. However, the iterative algorithm only needs a few or dozens of iterations to meet the convergence conditions for some linear systems, while it needs hundreds or even thousands of iterations for other linear systems. If the number of SpMV calls is small, using machine learning methods for prediction optimization may not yield performance benefits, as machine learning methods introduce additional overheads, including feature extraction, model inference, and format conversion. On the other hand, with the development of hardware devices and the demand for software applications, CPU-GPU has become one of the mainstream heterogeneous computing platforms. Software applications based on the CPU-GPU heterogeneous platform usually offload computing tasks to the GPU, while the CPU is always waiting for the GPU to complete tasks. However, modern servers are equipped with CPUs that typically have dozens of computing cores, supporting parallel computing with dozens or even hundreds of threads, and their computing power cannot be underestimated.

Based on the above considerations, this paper proposes an optimization method of cascaded prediction and asynchronous execution on heterogeneous platforms. In the asynchronous execution optimization, the traditional sequential execution of two steps is replaced by asynchronous execution of machine learning model prediction and iterative solver. Some threads are created on the CPU side to complete feature extraction and model inference, thereby hiding the additional overhead caused by machine learning optimization. In the cascaded prediction for different configurations, we do not need to wait for all configurations to complete prediction before updating the computational configuration in iteration, but can dynamically predict the optimal selection of each configuration in a cascading manner. Firstly, the optimal compression format recommended by the sparse matrix format prediction model is applied to the iterative solver. After the SpMV algorithm prediction model finds the optimal SpMV algorithm based on this format, the SpMV configuration in the iterative algorithm is updated. If further prediction of the parameter configuration of the SpMV kernel is required, the SpMV configuration in the iteration is updated after the prediction is completed.

\subsubsection{Cascaded Prediction}
\begin{figure}[!htbp]
    \centering
    \includegraphics[width=0.99\linewidth]{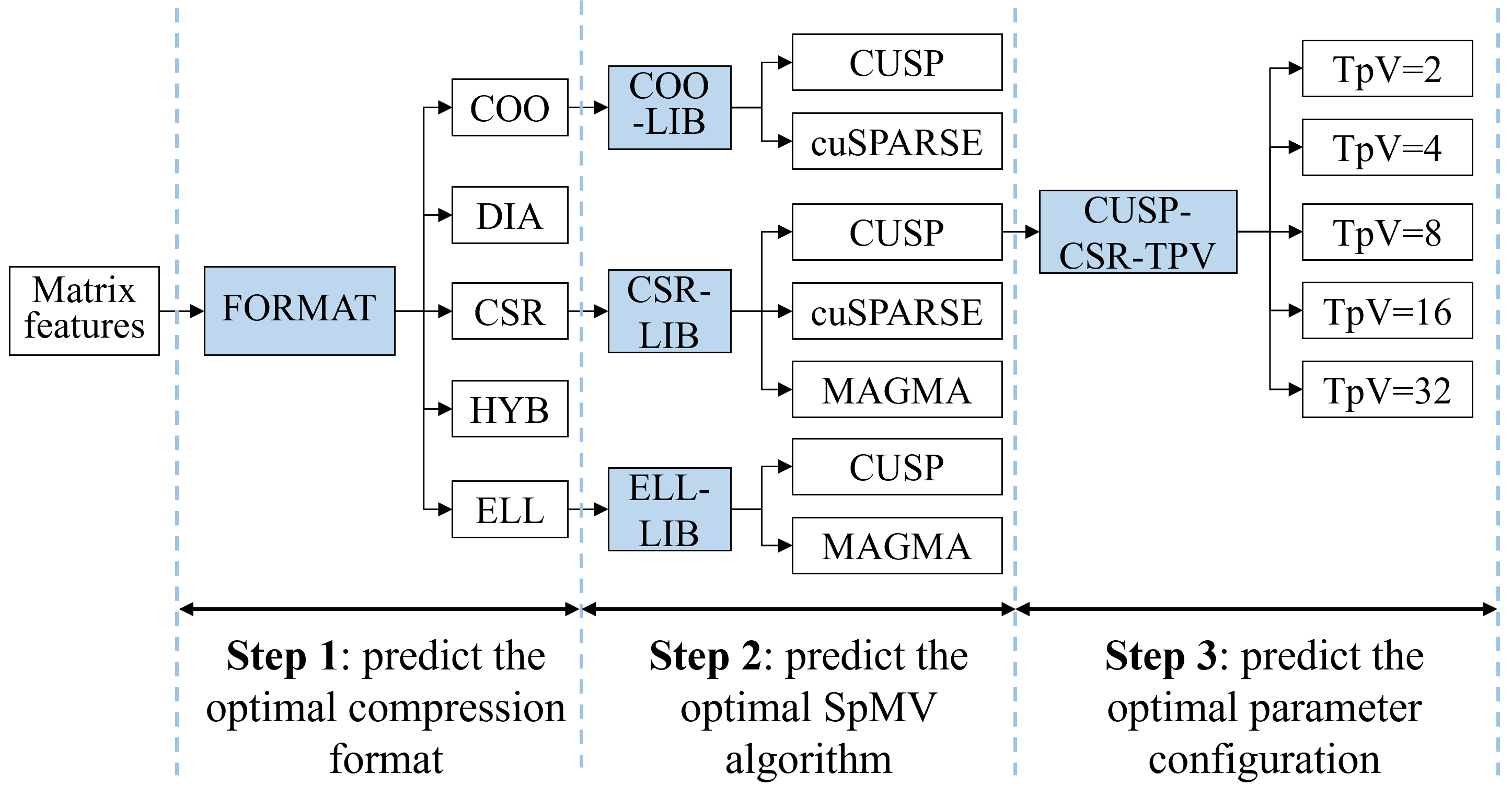}
    \caption{Cascaded prediction of different SpMV configurations.}
    \label{fig:cascade-format}
\end{figure}

Figure \ref{fig:cascade-format} presents the proposed cascaded prediction framework. First, the FORMAT model is used to predict the optimal compression format for the input matrix. If the prediction result is COO format, the COO-LIB model is used to predict the optimal SpMV algorithm in the computation library that supports COO-based SpMV calculations. If the optimal format predicted by the FORMAT model is CSR, the CSR-LIB model is used to select the optimal SpMV algorithm in the computation library that supports CSR-based SpMV calculation. If the CUSP library is selected, the CSR-CUSP-TPV model is further used to predict the optimal TpV configuration for the SpMV kernel. In summary, after the prediction of any model is completed, the SpMV called in the iterative algorithm can be updated accordingly in the next iteration.

When calculating configuration updates, there is usually a certain overhead associated. For example, if the predicted compression format is different from the default compression format, additional storage space needs to be allocated and the format needs to be converted. This paper chooses CUSP-COO as the default configuration to facilitate the conversion to other formats. In the CUSP library, the conversion of COO format to other formats can be completed by calling the corresponding CUDA kernel.
If the predicted computational library is different from the currently called library, it is necessary to switch the computational library. The required work is to extract the basic information of the matrix and the memory addresses of each array from the data structure representing the sparse matrix, and then load them into the data structure of the new computational library. This part of the overhead is usually very small and can be ignored. If it is necessary to change the configuration of TpV parameters, we only need to launch the SpMV kernel with the optimal thread configuration in the subsequent iteration, and the overhead can also be ignored. In summary, this paper provides a cascaded prediction method for multiple configurations, supporting the extension of more SpMV computational configurations.

\begin{figure}[!htbp]
    \centering
    \includegraphics[width=0.95\linewidth]{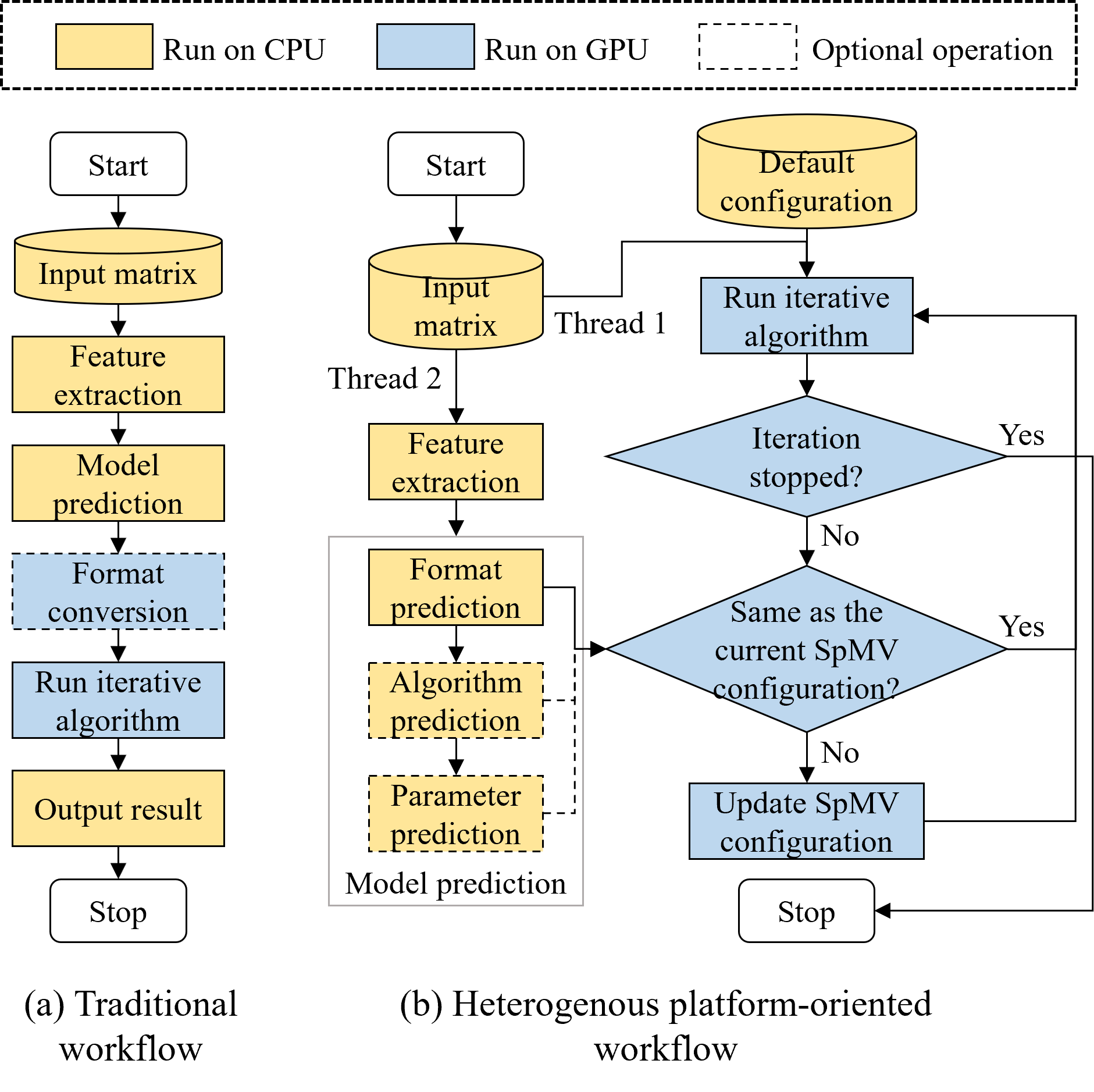}
    \caption{Runtime of iterative algorithms.}
    \label{fig:dynamic-ml}
\end{figure}

\subsubsection{Asynchronous Execution}
The asynchronous execution on the CPU-GPU heterogeneous platform is presented in Figure \ref{fig:dynamic-ml}. Figure \ref{fig:dynamic-ml}(a) presents the calculation process of the traditional iterative algorithm optimized using machine learning-based SpMV. Figure \ref{fig:dynamic-ml}(b) presents the calculation process optimized using asynchronous execution proposed in this paper. In the traditional method, given an input matrix, feature extraction and model prediction are performed sequentially. If the compression format needs to be changed, format conversion is performed. Then, the predicted SpMV configuration is used to run the iterative algorithm until the iteration terminates. Differently, in the method proposed in this paper, given an input matrix, some threads are initiated to control the CPU and GPU sides, respectively. The GPU side uses the default SpMV configuration for iterative computation. The default SpMV configuration selected in this paper is the COO-based SpMV implementation in CUSP. 
The CUSP computation library is selected because it provides an I/O interface for reading matrices and supports a wide range of sparse formats. At the same time, another CPU thread calculates the sparse matrix features required for model input and predicts the optimal choice for a certain configuration. If the prediction result differs from the current default configuration, the SpMV computation kernel called by the GPU is changed to use the SpMV kernel of the predicted configuration, and then iterative computation continues. Thread 2 listens for iterative computation on the GPU while performing feature computation and model predicting. If the iterative computation on the GPU ends before the completion of the feature calculation and model inference steps, the feature calculation or model inference is terminated.

In the aforementioned method, the configuration prediction on the CPU side and the iterative computation on the GPU side are performed concurrently. The computation configuration on the GPU is dynamically updated based on the prediction results from the CPU, hiding the overhead of feature computation and model prediction, and fully utilizing the computational resources of both the CPU and GPU.

\begin{table}[!htb]
    \centering
    \caption{Comparison of inference time between the original Python model and the C language code generated by m2cgen.}
    \scalebox{0.9}{
    \begin{tabular}{crrr}
        \hline
        \textbf{Model} & \textbf{Python (ms)} & \textbf{C (ms)} & \textbf{Speedup} \\
        \hline
        FORMAT       & 2.0513 & 0.0558 & 36.70 
\\
        COO-LIB      & 1.4091 & 0.0185 & 75.85 
\\
        CSR-LIB      & 1.8572 & 0.0025 & 729.40 
\\
        ELL-LIB      & 1.1451 & 0.0017 & 669.66 
\\
        CSR-CUSP-TPV & 2.3354 & 0.0019 & 1235.70 
 \\
        \hline
    \end{tabular}}
    \label{tab:m2cgen-vs-python}
\end{table}

In practical applications, machine learning models are typically trained and inferred using programs written in Python. Since Python is an interpreted language, its source code is interpreted line by line. This poses challenges to the efficiency of model inference, especially for SpMV, which has a millisecond-level running time.
In the method proposed in this paper, excessive model inference overhead may lead to the use of suboptimal SpMV configurations for a long time in iterative algorithms, missing a number of iterations. Given that the models selected in this paper are highly interpretable, they can be converted into compiled language programs with faster execution speeds, thereby enhancing the inference speed. Model-To-Code Generator (m2cgen) \cite{m2cgen} is a lightweight open-source tool used to convert trained ML models into code in various programming languages, including C, Java, Go, and C\#. 
Similar to \cite{jpdc/GaoJLWS24,10.1145/3664924}, this paper uses the tool to convert ML models into C language codes, allowing for seamless integration into iterative algorithms. Table \ref{tab:m2cgen-vs-python} shows the inference overhead of using the original Python ML model and the C language code generated by m2cgen in five classification tasks. It can be observed that the C language code achieves a significant speedup on six classifiers, with a maximum of 1,235x and an average of 549x. 

\begin{figure*}[!htbp]
    \centering
    \includegraphics[width=0.85\linewidth]{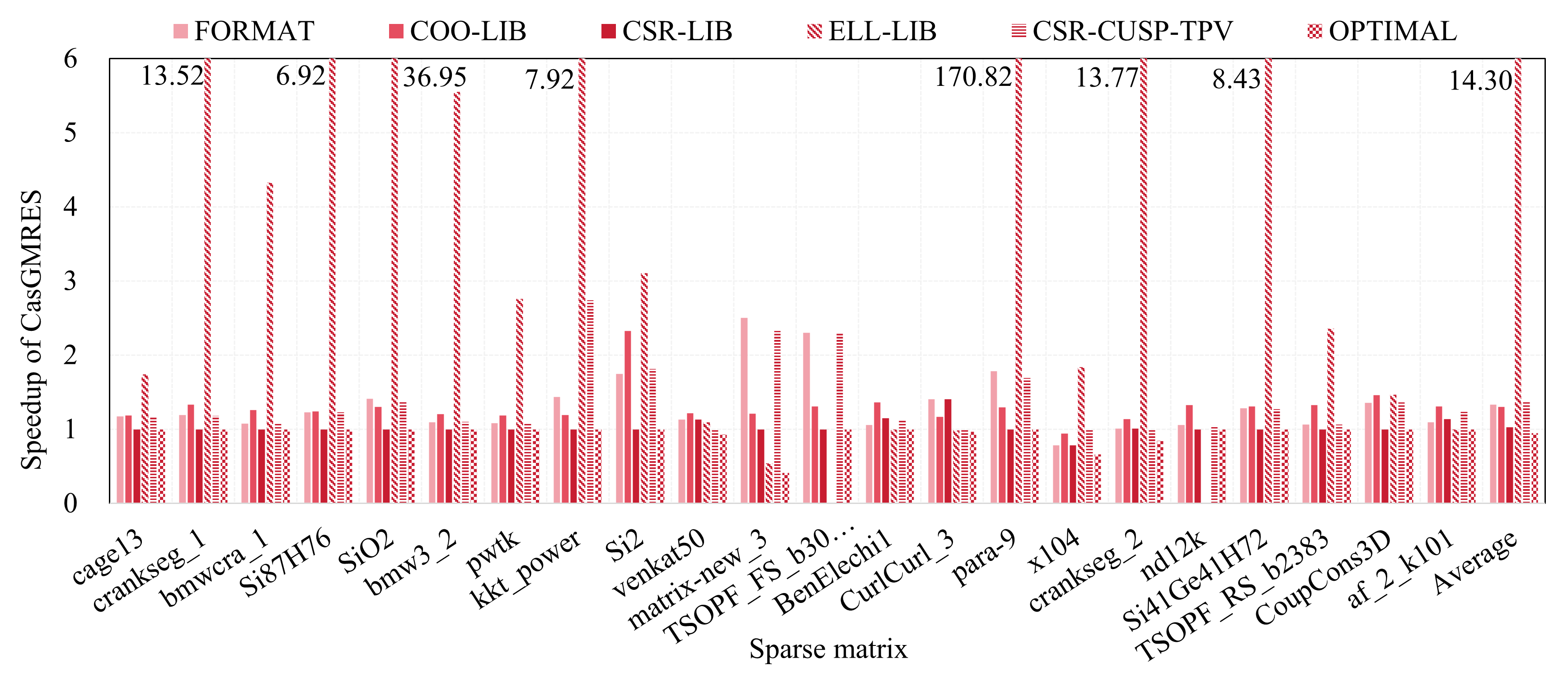}
    \caption{Performance speedup of SpMV using cascaded prediction (CasSpMV) against the SpMV using single area prediction (FORMAT, COO-LIB, CSR-LIB, ELL-LIB, CSR-CUSP-TPV) and the optimal SpMV configuration (OPTIMAL).}
    \label{fig:spmv-comp}
\end{figure*}


\section{Experimental Evaluation}
\subsection{Experimental Setup}
We conduct experiments on a computer equipped with both CPU and GPU devices. The CPU is Intel(R) Xeon(R) CPU E5-2680 v4, with a clock speed of 2.40 GHz, 56 cores, and 128 GB of main memory. The GPU is an NVIDIA Tesla V100, with 16 GB of device memory. The machine operates on Ubuntu 18.04.6 LTS, with GCC 8.4.0, CUDA 12.0, CUSP 0.6, and MAGMA 2.7.2, respectively.

The optimization method proposed in this paper is applicable to any iterative algorithm with SpMV as the performance bottleneck. In this paper, we take the GMRES iterative algorithm as an example to evaluate the performance of the proposed method. Our testing set includes 22 representative sparse matrices from the SuiteSparse Matrix Collection. Table \ref{tab:gmres-test-matrices} presents their rows, numbers of non-zeros, and corresponding GMRES iteration times.

\begin{table}[!htb]
    \centering
    \caption{Tested sparse matrix.}
    \scalebox{0.9}{
    \begin{tabular}{clrr}
        \toprule
        \textbf{ID} & \textbf{Matrix} & \textbf{Rows} & \textbf{NNZ} \\ \midrule
        1 & cage13 & 445,315 & 7,479,343\\
        2 & crankseg\_1 & 148,770 & 10,641,602\\
        3 & bmwcra\_1 & 148,770 & 10,641,602\\
        4 & Si87H76 & 240,369 & 10,661,631\\
        5 & SiO2 & 155,331 & 11,283,503\\
        6 & bmw3\_2 & 227,362 & 11,288,630 \\
        7 & pwtk & 217,918 & 11,524,432\\
        8 & kkt\_power & 2,063,494 & 12,771,361\\
        9 & Si2 & 769 & 17,801\\
        10 & venkat50 & 62,424 & 1,717,777\\
        11 & matrix-new\_3 & 125,329 & 893,984 \\
        12 & TSOPF\_FS\_b300\_c3 & 84,414 & 13,135,930\\
        13 & BenElechi1 & 245,874 & 13,150,496\\
        14 & CurlCurl\_3 & 1,219,574 & 13,544,618 \\
        15 & para-9 & 155,924 & 2,094,873\\
        16 & x104 & 108,384 & 8,713,602\\
        17 & crankseg\_2 & 63,838 & 14,148,858\\
        18 & nd12k & 36,000 & 14,220,946 \\
        19 & Si41Ge41H72 & 185,639 & 15,011,265\\
        20 & TSOPF\_RS\_b2833 & 38,120 & 16,171,169\\
        21 & CoupCons3D & 416,800 & 17,277,420\\
        22 & af\_2\_k101 & 503,625 & 17,550,675\\
        \bottomrule
    \end{tabular}}
    \label{tab:gmres-test-matrices}
\end{table}

\subsection{Cascaded Prediction Evaluation}

We first compare the SpMV performance achieved using the cascade or single-area prediction, with the results presented in Figure \ref{fig:spmv-comp}. 
A speedup of 1 indicates that the same SpMV configuration is selected by both methods. We can observe that our cascade prediction achieves superior or equivalent performance to single-area prediction on almost all matrices. For the ELL-LIB model, the unstable performance of the ELL format across different matrices results in the cascade prediction's SpMV performance significantly surpassing that of the ELL-LIB predicted on certain matrices. In summary, CasSpMV achieves an average speedup of 1.33x, 1.30x, 1.03x, 14.30x, and 1.37x relative to FORMAT, COO-LIB, CSR-LIB, ELL-LIB, and CSR-CUSP-TPV. Moreover, the optimal SpMV configuration is selected by CasSpMV for 17 sparse matrices.

\begin{figure}[!htbp]
    \centering
    \includegraphics[width=0.98\linewidth]{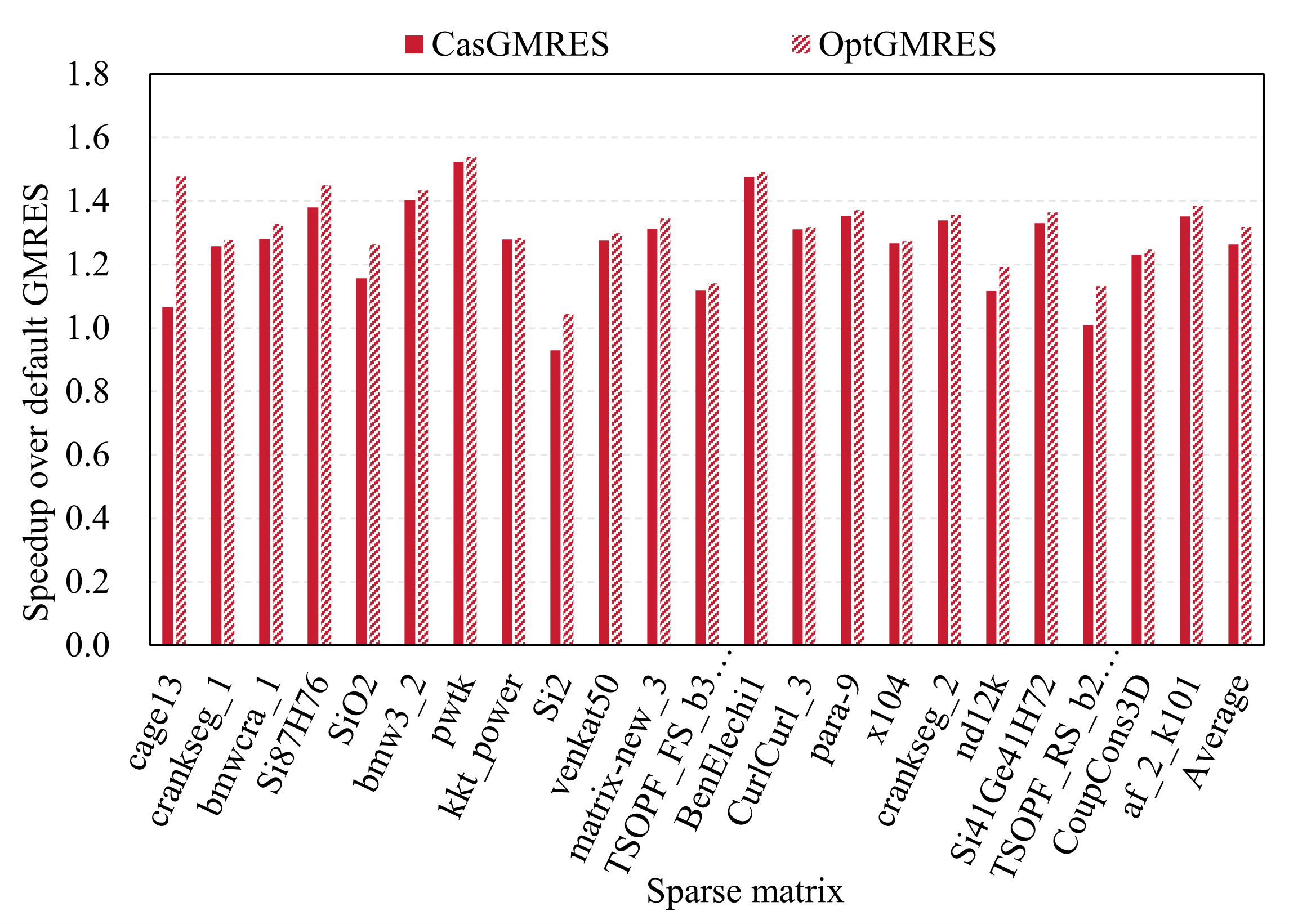}
    \caption{Speedup of GMRES using the cascaded predicted SpMV configuration (CasGMRES) and GMRES using optimal SpMV configuration (OptGMRES) relative to the GMRES using default SpMV configuration.}
    \label{fig:casgmres-vs-optgmres}
\end{figure}

Besides, we collect the GMRES running time using the CasSpMV and the GMRES running time using the optimal SpMV configuration, represented by CasGMRES and OptGMRES respectively. 
Figure \ref{fig:casgmres-vs-optgmres} presents their speedups relative to the GMRES with the default SpMV configuration. We can observe that CasGMRES achieves performance close to the optimal configuration on most matrices. Compared with the default SpMV configuration, CasGMRES achieves a maximum speedup of 1.52x and an average speedup of 1.26x. OptGMRES achieves a maximum speedup of 1.53x and an average speedup of 1.31x. 
Due to the fast convergence, matrix \textit{cage13} is always calculated using the default SpMV configuration, iterative computation at the GPU end is insensitive to cascading prediction at the CPU end. The speedup for \textit{TSOPF\_RS\_b2383} is close to 1 because its feature calculation is time-consuming, and the algorithm dose not switch to the predicted SpMV configuration until iteration 451. 
The speedup of CasGMRES is less than 1 for matrix \textit{Si2}, the reason is that the matrix is relatively small, and the performance gain brought by cascaded prediction cannot offset the cost of format conversion and ibrary switch. For the remaining 19 matrices, CasGMRES improves their GMRES performance, and the speedup is close to that of OptGMRES, indicating that the proposed cascaded prediction method enables the trained model to achieve high accuracy.

\subsection{Asynchronous Execution Evaluation}
\begin{figure}[!htbp]
    \centering
    \includegraphics[width=0.99\linewidth]{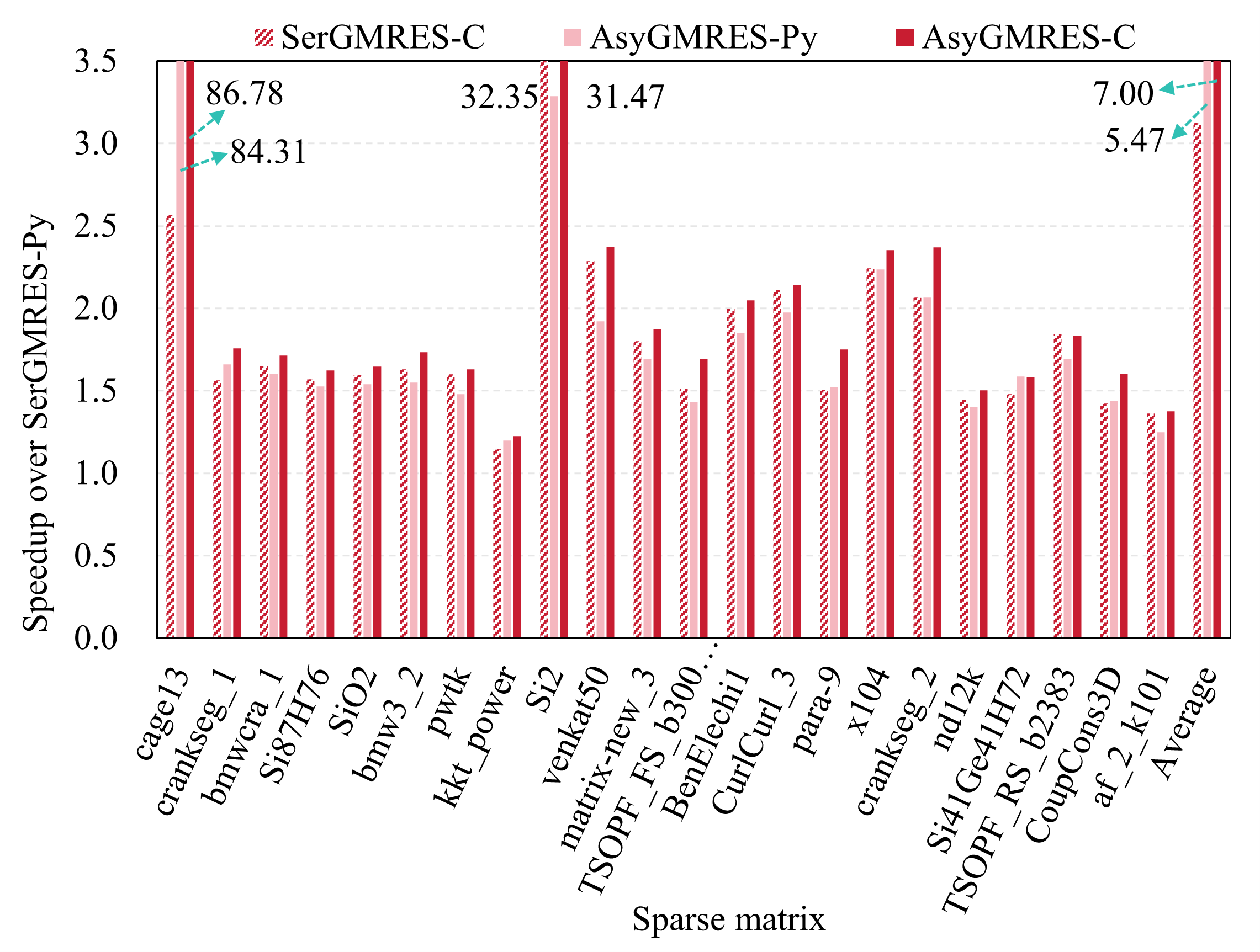}
    \caption{Performance comparison of GMRES using asynchronous execution (AsyGMRES-C, AsyGMRES-Py) and GMRES using sequential execution (SerGMRES-C, SerGMRES-Py). "C" and "Py" represent the C language codes and Python models are used for inference, respectively. SerGMRES-Py is the baseline. }
    \label{fig:casgmres-vs-sergmres}
\end{figure}

To evaluate the performance of the asynchronous execution, we developed the sequential version of cascaded prediction, referred to as SerGMRES, for comparison. SerGMRES first predicts the optimal SpMV and then applies the results to GMRES. Figure \ref{fig:casgmres-vs-sergmres} presents the performance comparison between asynchronous executed GMRES (AsyGMRES) and SerGMRES with the sequential and Python version GMRES. In the figure, "C" and "Py" indicate that the C language code and Python models are used for inference, respectively. 
We can observe that, except for \textit{Si2} and \textit{TSOPF\_RS\_b2383}, the speedup of AsyGMRES-C is always higher than that of SerGMRES-C. Overall, AsyGMRES-C and SerGMRES-C achieve average speedups of 7.00x and 3.13x respectively.
Both two methods achieve the highest speedup on \textit{cage13}, as it takes a short iteration running time and is greatly affected by the overhead of feature extraction and model inference.
For \textit{Si2} and \textit{TSOPF\_RS\_b2383}, the default configuration used by AsyGMRES-C in the previous iterations is not the optimal configuration, while SerGMRES-C uses the optimal SpMV configuration at the beginning of the iteration. 
In summary, AsyGMRES-C achieves an average speedup of 2.55x relative to SerGMRES-C, proving that the asynchronous execution of iteration algorithms on CPU-GPU heterogeneous platforms is effective.

\subsection{Inference Cost Evaluation}
We use the m2cgen tool to convert ML models into C language code, enabling the predicted optimal configuration to be applied to iterative algorithms as early as possible. Table \ref{tab:config-update-iter} compares the iteration time for various SpMV configuration updates in AsyGMRES-C and AsyGMRES-Py. It can be seen that, except for the matrix \textit{cage13}, where the convergence speed is faster than the prediction speed, AsyGMRES-C can complete the prediction and update of different SpMV configurations within about 3 iterations for most matrices. For \textit{para-9}, \textit{TSOPF\_RS\_b2833}, and \textit{CoupCons3D}, the prediction overhead is higher than for other matrices, but it does not exceed the prediction overhead of AsyGMRES-Py. For most matrices, AsyGMRES-Py requires more than 1,000 iterations to complete the final configuration updates, and in 7 matrices, the iteration ends but the SpMV configuration has not been predicted yet.

\begin{table}[!htb]
    \centering
    \caption{The number of iterations for updating SpMV configuration in AsyGMRES-C and AsyGMRES-Py.}
    \scalebox{0.8}{
    \begin{tabular}{lcccccc}
        \toprule
        \multirow{2}{*}{\textbf{Matrix}} & \multicolumn{3}{c}{\textbf{AsyGMRES-C}} & \multicolumn{3}{c}{\textbf{AsyGMRES-Py}} \\ \cmidrule{2-7}
        & FORMAT & LIB & TPV & FORMAT & LIB  & TPV \\ \midrule
        cage13 & $\times$      & -   & -   & $\times$      & -    & -   \\
        crankseg\_1 & 1      & 1   & -   & 401    & 1370 & -   \\
        bmwcra\_1 & 1      & 1   & -   & 451    & 1447 & -   \\
        Si87H76              & 1      & 1   & -   & 251    & 1281 & -   \\
        SiO2                 & 1      & 1   & -   & 401    & 1417 & -   \\
        bmw3\_2              & 1      & 1   & -   & 451    & 1420 & -   \\
        pwtk                 & 1      & 1   & -   & 301    & 1260 & -   \\
        kkt\_power           & 1      & 1   & -   & 101    & 447  & -   \\
        Si2                  & 1      & 2   & -   & $\times$      & -    & -   \\
        venkat50             & 1      & -   & 1   & 751    & -    & $\times$   \\
        matrix-new\_3        & 1      & 1   & -   & 751    & 1788 & -   \\
        TSOPF\_FS\_b300\_c3  & 1      & 1   & -   & 201    & 1187 & -   \\
        BenElechi1           & 1      & 1   & -   & 251    & 1251 & -   \\
        CurlCurl\_3          & 1      & -   & 1   & 151    & -    & $\times$   \\
        para-9               & 51     & 201 & -   & 1051   & $\times$    & -   \\
        x104                 & 1      & -   & 1   & 551    & -    & $\times$   \\
        crankseg\_2          & 1      & -   & 1   & 151    & -    & $\times$  \\
        nd12k                & 1      & 1   & -   & 101    & 948  & -   \\ 
        Si41Ge41H72          & 1      & 1   & -   & 251    & 1183 & -   \\
        OPF\_RS\_b2833       & 451    & 451 & -   & 901    & 1741 & -   \\
        CoupCons3D           & 151    & 151 & -   & 351    & 1098 & -   \\
        af\_2\_k101          & 1      & 1   & -   & 101    & 701  & -   \\
        \hline
    \end{tabular}}
    \label{tab:config-update-iter}
\end{table}

From Figure \ref{fig:casgmres-vs-sergmres}, we also can observe that, on almost all matrices, the speedup of AsyGMRES-C is always higher than that of AsyGMRES-Py, and the performance of SerGMRES-C is always higher than that of SerGMRES-Py. Overall, SerGMRES-C achieves an average speedup of 3.13x relative to SerGMRES-Py, and AsyGMRES-C achieves an average speedup of 1.48x against AsyGMRES-Py.

\section{Related Work}
\subsection{Sparse Matrix Compression Format and SpMV}
COO (Coordinate), CSR (Compressed Sparse Row), ELL (Ellpack), and DIA (Diagonal) are the four earliest basic compression formats \cite{paolini1989data}\cite{DBLP:journals/pc/Peters91}\cite{saad1994sparskit}. Subsequently, researchers have proposed some improvement strategies to compensate for the shortcomings of these basic formats, such as slicing \cite{kreutzer2014unified}\cite{europar/SunZWLZL11/CRSD} or blocking \cite{AshariSES14_BRC}\cite{tjs/LiXCLYLGGX20}\cite{Yan2014}\cite{NiuLDJ0T21_TileSpMV} based on the basic formats,
Mixing two or more basic formats, such as COO+ELL \cite{AnztCYDFNTTW20}\cite{bell2008efficient}\cite{BellG09}, CSR+ELL \cite{2016_AMB}\cite{jcss/YangLL18}, CSR+DIA \cite{DBLP:journals/ijhpca/YangLLSW14}\cite{DBLP:journals/tpds/GaoJTWS22}, more than two compression formats \cite{TAN2020521}\cite{NiuLDJ0T21_TileSpMV}. Designing new compression formats for the problem of uneven load caused by uneven distribution of non-zero elements in matrix rows, such as CSR5 \cite{CSR5}, CSR2 \cite{ccgrid/BianHDLW20}, CSRX \cite{ppopp/KourtisKGK11}, Cocktail \cite{cocktail2012}, or designing new SpMV algorithms, such as CSR-Scalar, CSR-Vector \cite{bell2008efficient}\cite{BellG09}, CSR-I \cite{DBLP:conf/europar/FlegarQ17}, CSR-Adaptive \cite{DBLP:conf/hipc/DagaG15}, AMF-CSR \cite{DBLP:conf/icpads/GaoJLSWS21}, LightSpMV \cite{liu2018lightspmv}, merge-based SpMV \cite{MergedSpMV_SC16}. Given the widespread application of sparse matrices and the diversity of compressed storage formats, academia, and industry have developed multiple sparse matrix computation libraries to support applications related to sparse matrix computations, such as cuSPARSE \cite{nvidia2023}, Eigen \cite{guennebaud2010eigen}\cite{eigen2013}\cite{CUSP}, MKL \cite{intel2023mkl}, MAGMA \cite{magma-library}, PETSc \cite{balay2019petsc}, Ginkgo \cite{Ginkgo}, and rocSPARSE \cite{rocSparse}.

\subsection{Machine Learning-based SpMV Optimization}
Regardless of the compression format, SpMV algorithm, or parameter settings in the SpMV kernel, there is no single SpMV configuration that can achieve optimal performance on any matrix due to the diverse non-zero element distributions of sparse matrices. Therefore, researchers used ML models to model the relationship between sparse matrices and optimal SpMV configuration. Armstrong et al. \cite{Armstrong2008_RL} used reinforcement learning techniques to select the optimal compression format. The optimal format prediction models of Li et al. \cite{SMAT2013}, Sedaghati et al. \cite{DBLP:conf/ics/SedaghatiMPPS15}, and Chen et al. \cite{Chen2018} are all based on decision trees. Akrem et al. \cite{Akrem2016ICPP_SVM} used a multi-class support vector machine (SVM) classifier to select the optimal format, and in \cite{icpads/BenatiaJWS16} used support vector regression (SVR) and multilayer perceptron (MLP) to predict the SpMV running time. In the later proposed BestSF (Best Sparse Format) \cite{Akrem-BestSF}, a weighted SVM model was used. Other works that use ML methods to predict the optimal compression format include \cite{Mehrez2018HPCS_SVM}\cite{Nisa2018IPDPSW}\cite{ShenXipeng2018PPoPP_bridgingGap}\cite{DBLP:journals/ppl/Hamdi-LarbiMD21}, and \cite{Shenxipeng2020TPDS_overheadConscious}.

In addition to predicting the optimal format, some work focuses on predicting the optimal SpMV algorithm. For example, Hou et al. \cite{2017-IPDPSW-autoTuning} used a decision tree based model to select the optimal kernel from three CSR-based SpMV kernels. He et al. \cite{DBLP:journals/ieicet/CuiHKT18} used feature maps instead of hand-selected features. Dufrechou et al. \cite{ijhpca/DufrechouEQ21} used a bagged tree-based model to predict the SpMV implementation with the lowest energy consumption. 
The classifier proposed by Xiao et al. \cite{DBLP:conf/hpcc/XiaoZCHL22} considers the changes in the sparsity of the multiplied vectors during the iterative process of graph analysis applications. Besides, some researchers use ML models to predict the execution time of SpMV kernels \cite{Akrem2020}\cite{ipps/NisaSSVS18}\cite{barreda2020performance}\cite{barreda2021convolutional}, while some others use ML model to predict the optimal parameter settings, such as predicting the optimal number of processes and the optimal mapping of data on the underlying hardware \cite{usman2019zaki} \cite{DBLP:journals/access/UsmanMKA19}, the optimal block size in BCSR \cite{ahmed2022aaqal}, and the optimal thread configuration in the CSR-Vector algorithm \cite{jpdc/GaoJLWS24}. Further exploration of SpMV can be found within a dedicated survey paper \cite{SPMV-survey}.

\subsection{Other Work}
The overhead introduced by preprocessing may offset the performance benefits brought by SpMV optimization. Therefore, Shen et al. \cite{Shenxipeng2020TPDS_overheadConscious} proposed an overhead-conscious SpMV compression format selection method for iterative algorithms. It uses the default SpMV kernel in the initial iterations and then predicts the total number of iterations based on the initial iterations's residuals. If the predicted iteration count is greater than a certain threshold, ML-based format prediction is used, and SpMV computation is performed based on the predicted format. Compared with our work, this method introduces the overhead of predicting the iteration count in the first stage, and for cases where the predicted iteration count is greater than the threshold, it does not hide the overhead of feature extraction, model prediction, and format conversion.

Swann et al. \cite{cgo/SwannOSM24} explored the selection of runtime kernels for irregular problems, mainly focusing on reducing the cost of feature extraction. They divided the features into two categories: known features and gathered features. Known features are easily obtained, and gathered features have higher computational complexity. They trained a lightweight decision tree based classifier to decide whether it is necessary to collect gathered features. If not, a predictor based on known features is used to predict the optimal SpMV kernel; otherwise, the feature extraction kernel is executed, and a predictor corresponding to complex features is used to predict the optimal SpMV kernel. This method avoids collecting complex features on some problems through a lightweight DT-based classifier, but the cost of model prediction still exists. Moreover, if it is necessary to collect complex features, the costs of feature extraction and model prediction still exist.

\section{Conclusion}
This paper introduces cascaded SpMV configuration prediction and asynchronous execution optimization for iterative algorithms on heterogeneous platforms. 
To address the suboptimal SpMV performance of a single SpMV configuration prediction, we propose a cascaded prediction approach for various SpMV configurations, recommending the optimal configuration starting with the sparse compression format, followed by the SpMV algorithm and thread configuration. To counterbalance the preprocessing overhead against the performance benefits of SpMV, we present an asynchronous execution model. While the iterative algorithm using the default SpMV configuration operates on the GPU, the CPU side executes matrix feature extraction and machine learning model inference. Experimental evaluation on 22 sparse matrices demonstrates that the SpMV using cascaded prediction achieves an average speedup of 1.33x relative to the SpMV using only compression format prediction. Moreover, the iterative algorithm using asynchronous execution and cascaded prediction optimization achieves an average speedup of 2.55x relative to the optimized sequential prediction version.

\bibliographystyle{IEEEtran}
\bibliography{IEEEabrv,ref.bib}

\end{document}